\documentclass[aps,pra,twocolumn,superscriptaddress,showpacs]{revtex4-1}
\usepackage[english]{babel}
\usepackage{epsfig}
\usepackage{epstopdf}
\usepackage{graphicx}
\usepackage[utf8]{inputenc}
\usepackage{verbatim}
\usepackage{float}
\usepackage{amsmath}
\usepackage{amssymb}
\usepackage{color}
\usepackage{url}

\usepackage{datetime}
\def\be{\begin{equation}}
\def\ee{\end{equation}}
\def\e{{\rm e}}
\def\d{{\rm d}}
\def\({\left(}
\def\){\right)}
\begin{document}
\title{\Large The C$_{60}^-$ thermal electron emission rate}
\author{K. Hansen}\thanks{klavshansen@tju.edu.cn}
\affiliation{Center for Joint Quantum Studies and Department of Physics, 
School of Science, Tianjin University, 92 Weijin Road, Tianjin 300072, China}
\date{\today,~\currenttime}
\begin{abstract}
The thermal electron emission rate constant for C$_{60}^-$ has been 
deduced for internal energies from 9.7 to 14.1 eV
from storage ring measurements 
of the decays of ions reheated with single photons absorption.
The thermal radiation from the ions is quantified from the data
with respect to 
continuous cooling and discrete photon quenching.
\end{abstract}
\maketitle

\section{Introduction}

Measurements of rate constants in molecular beams with standard 
approaches require very good control over the excitation energy. 
A width in the internal energy distribution in a decaying particle
or molecule will also introduce 
a width in the distribution of rate constants in the beam molecules, 
and due to the strong dependence of rate constants on excitation 
energy, any spread in energy is strongly amplified for the rate 
constants.
This makes direct measurement of rate constants very difficult even 
for  relatively narrow internal energy distributions.
The problem is not solved by extracting molecules from canonical 
thermal distributions into molecular beams, as demonstrated in 
\cite{andersen2002thermionic} with a calculation of a numerical 
example for C$_{60}^-$.

The detrimental consequence of a finite width of the 
internal energy distribution to a simpleminded determination of
rate constants is perhaps best
demonstrated by considering the measured decay rates vs. the 
rate constants present in the ensemble.
Allowing for measurements over all possible time scales, 
an Arrhenius-type analysis would be based on the approximation
\be
k(\langle E \rangle) \stackrel{?}{\approx} \langle k(E) \rangle,
\ee
where the average is over the energy distribution. 
This is emphatically wrong. 
Adding a finite dynamic time range to the measurement of 
the right hand side will make this poor approximation even 
worse.

In fact, when the energy distribution of the 
molecules in a beam is 
sufficiently broad and in the absence of competing channels, 
molecular decay will occur with a rate with a time dependence 
close to $1/t$ \cite{hansen2001}. 
In the presence of the frequently occurring phenomenon of 
thermal radiation, this power law will be suppressed 
at long times with an almost exponential time dependence 
\cite{HansenRMS2020}.

Situations with broad energy distributions 
are seen for all particle sizes but arise particular 
frequently for large molecules and clusters, because for these,
excitation to internal energies where 
reactions occur on measurable time scales will require large 
amounts of energy, up to several tens of eV. 
Deposition of precise amounts of energies of such magnitudes is 
a very challenging experimental task.
Photo excitation experiments with a single high energy photon, 
for example, will often lead to direct (first or secondary) 
ionization of the molecules or to electron detachment from anions. 
The alternative strategy of multiple absorption of smaller energy 
photons suffers from the inherent spreads in absorption statistics.
Collisional excitation is possible, as demonstrated with electron 
collisions with fullerenes \cite{MattZPD1997,MattPTRSL1999}, 
but these suffer from a broad energy transfer efficiency, 
requiring a detailed quantitative analysis of the reaction products 
with a number of highly non-trivial assumptions.

The origin of the power law behavior is the loss of a well defined 
energy scale in the excitation energy distribution caused either by
such post-production excitation or by the use of hot sources, which
almost unavoidably produce clusters with broad energy distributions. 
For a unimolecular decay in vacuum, loss of an energy scale is 
equivalent to loss of a time scale. 
This is reflected in the absence of a characteristic time scale in 
the $1/t$ dependence of the decay rate.

If one wants to measure absolute energies under such 
conditions, it is therefore necessary to introduce an energy scale
by hand.
Doing so, it turns out that such a procedure allows
to determine the rate constant also for these broad energy 
energy distributions.
The demonstration of this procedure is the purpose of this article.

The reheating experiments that provide the
data for the analysis here were obtained with
C$_{60}^-$ in the experiments reported in 
\cite{SundenPRL2009}. 
In these experiments, the anions were created 
hot from the source and injected into an electrostatic storage 
ring, where they decayed by spontaneous electron emission. 
At a variable time after production, a small 
fraction of the un-decayed ions were reheated by one-photon 
excitation.
The photon energy absorbed and dissipated caused a heating of 
the molecule that lead to an enhanced delayed thermal 
electron emission. 
The time profile of the enhanced decay was used to locate the 
equivalent backshifted time, i.e. the time where the spontaneously 
decaying ions decayed with the same time dependence as the laser 
excited ions. 
An overall multiplicative constant on the enhanced decay, 
which reflects quantities such as laser fluence, beam overlap, 
and photon absorption cross section was only relevant for the 
amplitude of the laser enhanced signal and did not enter the 
analysis.

Together with the instrumental laser firing time, the determination 
of this apparent shift of the zero time of the power law decay 
due to the reheating provides the time interval during which one 
photon energy was lost.
The procedure can therefore be used to determine the absolute cooling 
rate of the ions.
The results obtained were in very good agreement with the known 
facts of C$_{60}^-$, such as the electron affinity and also with 
the model for the radiative cooling developed in 
\cite{AndersenEPJD2000}.

The data from this experiment somewhat surprisingly also allow
for the determination of the parameters that determine the 
energy resolved rate constants. 
Furthermore, they provide a measure of the relative importance
of continuous and discrete cooling. These two types of thermal 
photon emission differ only by the magnitude of 
the energies of the photons emitted, and thereby by the effect 
they have on the measured decay dynamics of the ions.

The outcome of the analysis of the C$_{60}^-$ data will provide 
the absolute decay rate, parametrized by the product of activation 
energy and heat capacity, the frequency factor of the rate constant, 
and a binary spectral distribution of the thermally emitted 
photons, all pertaining to the excitation energy 
interval between 9.7 and 14.2 eV.

The remainder of the paper is divided into a section where the 
theory behind the experimental data and the present analysis is 
described in some detail. This is followed by a section where the 
experiments are described, after which a section presents the 
data analysis and the results.
Finally, the procedure and the results are summarized and discussed.

\section{Theoretical background}

The spontaneous statistical decay rate 
of an ensemble of particles 
in a molecular beam is given by the decay rate 
constant averaged over all excitation energies present 
in the ensemble \cite{andersen2002thermionic,HansenRMS2020};
\be
\label{rate1}
R(t) \propto \int_0^{\infty} g(E) k(E) \exp(-k(E)t) {\d}E.
\ee
where $R$ is the measured decay rate, i.e. the number of 
decays per time unit, and $k(E)$ is the decay rate constant 
of an ion with excitation energy $E$. 
The decay rate 
on the left hand side of Eq.(\ref{rate1})
is not any proxy for a rate constant, it should be emphasized. 
Decays of systems with a specific energy remain exponential, as the
equation also assumes.
The quantity $g(E)$ is the ensemble density of excitation energy 
at the time it was created in the source, and $t$ is the time elapsed 
from the creation of $g$ in the source to the measurement.
The constant of proportionality is the combined transmission and 
detection efficiency. 
When $g(E)$ is broad, the integrand peaks at the rate constant 
$k_m$ for which 
\be
\label{maxk}
\frac{{\d}}{{\d} E} k_m(E) \exp(-k_m(E)t) =0 \Rightarrow k_m(E) t = 1,
\ee
corresponding to a peak value of $\exp(-1)/t$ for $k(E) \exp(-k(E)t)$.
This result is derived without specifying the expression for $k(E)$
and holds generally, insofar as Eq.(\ref{maxk}) has solutions,
which may not be the case for ultrafast processes but will be the case 
for measurement times relevant here. 
Implicit in the calculation is the assumption that
the decay process involves an activation energy, and the obvious 
requirement that it is observable, i.e. leads to a change in 
mass or charge state.
The equation only has one solution
if $k(E)$ is a monotonically increasing function of $E$, which can 
also be safely assumed here, but which might not be true
around a phase transition.

\begin{figure}[ht]
\includegraphics[width=0.3\textwidth,angle=270]{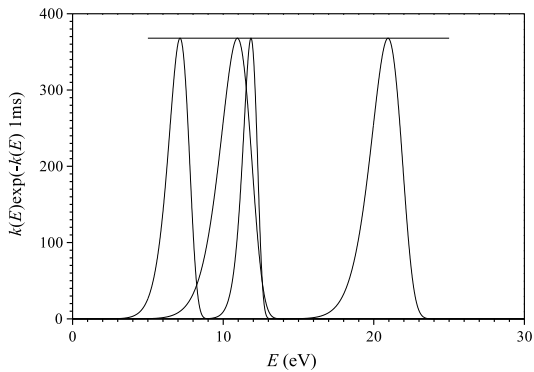}
\begin{centering}
\caption{The product of rate constants and survival probability 
for a broad energy distribution after
1 ms for a few different parameters for the rate constant given 
in the main text.
The line is the calculated value of $1000/\exp(1)$ s$^{-1}$
The parameters are, from low to high peak energies: 
$(\omega,\phi C_v, E')=(10^{14}~{\rm Hz},434 ~{\rm eV},10~ {\rm eV})$,
$(10^{12}{\rm ~Hz},434~ {\rm eV}, 10 ~{\rm eV})$, $(10^{14}{~\rm Hz},
300 ~{\rm eV},0)$, and $(10^{12}{~\rm Hz},434 ~{\rm eV}, 0)$.
\label{peakK}}
\end{centering}
\end{figure}

It is worth demonstrating the generality of the result in 
Eq.(\ref{maxk}) with some different expressions for rate constants.
Fig.(\ref{peakK}) shows a few examples. 
The expression for the decay constant, which will also be used in 
the analysis, is of the simple form
\be
\label{kstand}
k = \omega \exp \( -\frac{\phi C_v}{E+E'}\).
\ee 
Here $C_v$ is the canonical heat capacity in units of $k_B$, less one
due to the microcanonical correction \cite{andersen2001}
($k_B=1$ will be used throughout), and $\phi$ is the decay channel 
activation energy. The parameter $\omega$ plays the 
same role as the frequency factor in the canonical Arrhenius  
expression, although the two numerical values are in general different.
The energy $E'$ is the offset in the caloric curve,
$E = C_vT + E'$, which can not be assumed zero.
For thermionic emission from C$_{60}^-$ the parameter $\phi$
is to a first 
approximation expected to be the electron affinity of 2.67-2.68 eV 
\cite{BrinkCPL1995,HuangJCP2014}. 
In spite of its simplicity, Eq.(\ref{kstand}) is very accurate for 
our purpose because only an energy interval of ca. 4 eV is 
covered in the experiments here. This question is discussed in the 
appendix and corrections to parameters made in the discussion 
section.

The rate constants used in Fig.(\ref{peakK}) are all variations 
of the rate constant in Eq.(\ref{kstand}).
Other examples with different functional forms are given in 
\cite{HansenRMS2020}, with an identical 
conclusion concerning the peak values.

The decay rate is the integral of the peaks in Fig.(\ref{peakK}).
The integral is on the order of the width times the 
height. 
The order of magnitude of the widths of the peaks, 
$\delta E$, is given by
\be
\frac{\d \ln\({\rm e}^{-kt}\)}{\d E} \delta E \sim 1 \Rightarrow 
\delta E \sim \(\frac{\d \ln k}{\d E}\)^{-1},
\ee
where $k=1/t$ was used.
Given the rapid variation of $k$ with energy, this will remain
fairly constant over a wide range of times. 
This suggests the possibility that also the decay rate
may vary approximately as $1/t$.
To examine this question the decay rate is
calculated by considering the time 
dependence of the energy distributions.
Ignoring the variation of $g$ with energy, the energy distribution 
of the surviving ions, $\exp(-k(E)t)$, is essentially constant up 
to an energy close to the value defined by $k(E)=1/t$, at 
which point it bends over and rapidly approaches zero. 
The motion of this cross-over energy with time represents the decay 
rate.
Solving Eq.(\ref{kstand}) for $E$ and differentiating
with respect to time then gives the decay rate
\be
R(t) = -c'g\frac{\d E(k=1/t)}{\d t} = 
c'g\frac{\phi C_v}{(\ln(\omega t))^2} \frac{1}{t},
\ee
where $c'$ is a constant that includes the detection, 
transmission efficiency and other instrumental parameters, and 
$g(E)$ is set to a constant, $g = g(E(k=1/t))$. 
Absorbing $g$ into the constant, $c \equiv c'g$, and rewriting gives
\be
\label{kpeak}
\frac{1}{t} = k_m =  
R(t) \frac{(\ln(\omega t))^2}{c\phi C_v},
\ee
where $k_m$ is the value for which the decay peaks.
The difference between the time dependence of the decay rate and 
the rate constant at peak decay rate is the time variation of the 
width of the decaying peak considered a function of excitation 
energy, and is summarized by the factor $(\ln(\omega t))^2$.
This equation has been established previously, see e.g.
\cite{HansenRMS2020} and references therein, and has been used
to determine heat capacities of water clusters and radiative 
cooling of cationic carbon clusters experimentally, for example.

In the presence of thermal radiation, relevant for 
C$_{60}^-$ in the experimental data used here, the relation must 
be reconsidered.
In principle also the C$_2$ emission is a possible channel. 
However, this has an activation energy which is close to four times
that of electron emission from the anion and can safely be ignored.
The only channel competing with electron emission is therefore 
thermal radiation.

In the context of ensembles there are two categories of thermal 
radiation, defined by the magnitude of the energies of the 
emitted photons.
When the emission is by sufficiently low energy photons, the 
radiation is effectively a continuous cooling. 
This means that the energy distribution shifts down with time 
similarly to the non-radiative situation, just faster.
The shape of the cross-over region of the energy distribution is 
virtually unchanged in this small photon energy limit. 
When only this type of radiation is present, its effect can be 
determined from the observed decay rate with an expression 
analogous to Eq.(\ref{kpeak}) where $t'$ is given by
\be
\label{kpeak-rad}
R(t) = \frac{c g\phi C_v}{t'\ln(\omega t')^2},
\ee
from which the peak rate constant is identified as
\be
k(t) = \frac{1}{t'}
\ee
where $t'$ is a fictitious time which is equal to the time needed 
to wait to have an identical decay rate in the absence of radiation.
The decay at short times which is not influenced by any radiative 
cooling can be used to determine the constant 
of proportionality that links $t$ and $t'$. 
In the logarithm the difference between the physical time and 
$t'$ can often be ignored. 

When large energy photons are emitted, the simple powerlaw relation 
needs to be modified once more. 
Photon energies are considered large if the emission of a single
photon will quench the decay on a time scale corresponding to the 
rate constant after emission. 
The precise energy where this shift from continuous cooling to 
quenching happens was analyzed in \cite{FerrariIRPC2019}, and will 
be discussed here after the presence of these photons is quantified. 

For the fullerenes, the largest part of the radiation is well 
understood as being carried by the broad surface plasmon resonance 
\cite{AndersenEPJD2000}.
Although centered at 20 eV, it reaches into the near infrared which 
allows the low energy tail to be excited thermally with an oscillator 
strength which gives a radiative energy emission rate which is two 
orders of magnitude higher than the contribution from the 
vibrational transitions \cite{andersen1996}.
The calculated magnitude is consistent with both the anion 
cooling and the original observation of the strong radiative 
cooling of the much hotter fullerene cation fragments \cite{hansen1996}.
The distribution of photon energies generated by the plasmon 
resonance emission covers both the small and large values, and both 
types of channels therefore need to be considered in the analysis.

Whereas for small photon energies the emitted power is the relevant 
quantity, for large photon energies it is the emission rate constant. 
Although photon emission rate constants vary 
with the temperature to the power 6 \cite{andersen1996}, this is still 
slow compared to the variation of rate constant of the observed 
thermionic emission rate constant, and we can 
here set the discrete energy emission rate constant to a single 
value, $k_p$. Interestingly, the power of 6 which
originates in an photon absorption cross section that varies 
with the square of the photon energy, has also been observed for
larger, metallic nanoparticles \cite{GranqvistPRL1976,KimPRB1989}.
The presence of high energy photon radiation
means that the abundances at the decaying edge, 
and hence also the decay rates, are reduced by the factor $\exp(-k_p t)$.
Together with the effect of the continuous cooling and after
normalization to the short time behavior of $1/t$, the observed 
rate $R_n$ is then equal to 
\be
R_n(t) = \frac{1}{t'}\e^{-k_p t} = k(t)\e^{-k_p t},
\ee
or
\be
\label{kcorr}
k(t) = R_n(t){\e}^{k_p t},
\ee
where $k(t)$ is the thermionic rate constant at
the peak of the energy distribution.
The fitted curve from the experimentally measured spontaneous 
decay rate of C$_{60}^-$ from the hot source used 
gives the function \cite{SundenPRL2009}
\be
R_n(t) = \frac{1}{t} \exp\(-122{\rm s^{-1}}t+1320{\rm s^{-2}} t^2)\),
\ee
and hence
\be
\label{kcorr2}
k(t) = \frac{{\e}^{k_p t}}{t} \exp\(-122{\rm s^{-1}}t+1320{\rm s^{-2}} t^2)\).
\ee

\begin{figure}[ht]
\includegraphics[width=0.3\textwidth,angle=90]{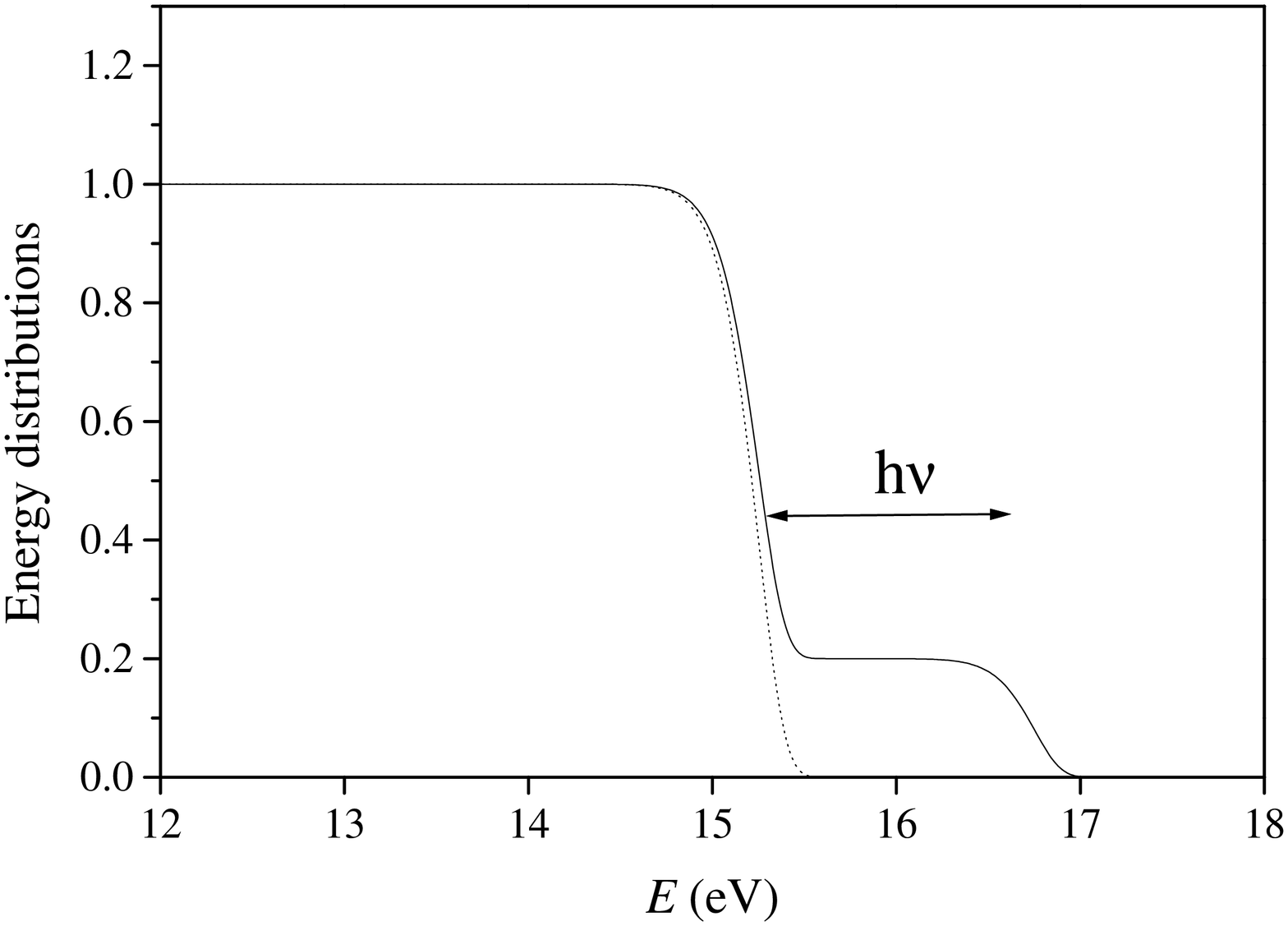}
\begin{centering}
\caption{A schematic view of the energy distributions immediately 
before (dotted line) and after (full line) a photon with energy 
$h\nu$ has been absorbed at $t_{las}$.\label{schematic}}
\end{centering}
\end{figure}

The analysis so far has only dealt with the spontaneous decay. 
If the molecule is exposed to a laser pulse some time after 
production, the absorbing fraction of the energy distribution 
will be shifted up by the photon energy. 
The situation is illustrated schematically in Fig.(\ref{schematic}).
The small fraction of the distribution that has been shifted up 
in energy has almost the same shape as the unshifted distribution 
had at some earlier time, apart from the absolute height.
This has been shown in \cite{HansenIJMS2018} to which the reader 
is referred for details of the calculation.
After photon absorption at $t_{las}$ the decay rate is therefore 
given by
\be
\label{Rbackshift}
R_{las}(t) = p\e^{-k_p \(t_{las}-t_0\)} R(t-t_{las}+t_0) + (1-p)R(t),
\ee
where $t_0$ is a backshifted time, and $p$ is the photon 
absorption probability. 
The spontaneous decay rate is a function of time and the 
physical interpretation of $t_0$ is that it is the time where the decay 
rate was equal to the rate observed after photon absorption.
This time can therefore be determined by a fit 
of the first term on the right hand side of Eq.(\ref{Rbackshift}) 
to the decay rate at earlier times. 
The fraction of absorbing ions, $p$, was so low in the experiments 
that is practically unobservable in the second term in the equation. 
This facilitated the analysis although it is not an essential 
requirement. 
The non-zero value of $k_p$ has no effect on this part of the 
analysis. 
It was not explicitly considered in \cite{SundenPRL2009}, 
but the cooling rates obtained there remain unchanged, although it 
is clear that they only refer to the small photon energy cooling
power.

The non-exponential decay is essential to determine the cooling
with this procedure because for non-exponential decays the value 
of both $p$ (or more precisely $p\e^{-k_p(t_{las}-t_0)}$) and 
$t_0$ can both be determined, a possibility which is not present 
for an exponential decay.

As shown, decay rates are proportional to decay constants and  
Eq.(\ref{Rbackshift}) therefore also holds for the peak 
distribution values $k_m(t)$ with the substitution $R(t)\rightarrow 
k_m(t)\e^{-k_pt}$.
The values of $t_0$ depend on the photon energy and $t_{las}$ 
but are independent of the absorption cross section and 
instrumental parameters.
Keeping the laser firing time $t_{las}$ fixed and varying the 
photon energy, it is therefore possible to obtain the variation 
of the rate constant with photon energy as
\be
\label{km1}
k_m(E(t_{las})+h\nu) = {\e}^{k_p t_0} R_n\(t_0(t_{las},h\nu)\),
\ee
and similarly 
\be
\label{km2}
k_m(E(t_{las})) = {\e}^{k_p t_{las}} R_n \( t_{las}\).
\ee
The energy $E(t_{las})$ is the energy where the decay rate 
peaks at time $t_{las}$. It is unknown but both $h\nu$ and 
$t_0$ are known. As indicated in Eq.(\ref{km1}), 
the measured values of $t_0$ depend on both the laser firing 
time and the photon energy. 
When considering decay rates in the following, the
term energy will always refer to this particular energy or the 
corresponding peak rate energy for the shifted distributions. 
In statements about the rate constant, the energy will refer to 
the argument in Eq.(\ref{kstand}).
Eqs.(\ref{km1},\ref{km2}) are the basic equations for the analysis
of the experimental data. They will be used below to 
express the $k$'s in terms of known, experimental times 
and parameters of the decay. 
The subscript $m$ indicates that the rate constant refers 
to a specific energy here. 
It will be dropped below.

It should be noted that although a number of measured 
values of $t_0$ correspond to times before the 
mass selection has been completed in these experiments, 
this causes no problem for the analysis, because other 
experiments on C$_{60}^-$ have established the short time
behavior as a well behaved power law, see e.g. 
\cite{andersen1996}, and this behavior is also
well established as a general phenomenon, see e.g. 
the examples listed in \cite{HansenRMS2020}.

\section{Experiments}

The data used for the analysis were recorded at the Tokyo 
Metropolitan electrostatic storage ring, TMU e-ring. 
The analysis of the absolute cooling rates derived from 
these data was published in \cite{SundenPRL2009}, and the 
description of the experiment here will be limited to the
pertinent points. For a detailed description of electrostatic 
storage rings and their use for decay measurements, the reader 
is referred to the rich literature on the subject, see e.g.  
\cite{SPMollerNIMA1997,JinnoNIMB2004,storagerings,
martin2013,SchmidtRSI2013,NakanoRSI2017}.

The C$_{60}^-$ anions were produced in a laser ablation source
without any cooling gas and injected into the ring together with 
some amount of other anionic carbon species produced during the 
ablation, mainly other fullerenes. 
The circulation time of C$_{60}^-$ in the ring was 122 $\mu s$.
A set of pulsed deflection plates was used to eject the unwanted 
species, based on their mass dependent circulation time. 
This beam purification process was completed within 1 ms after 
production of the ions in the source.

After a variable storage time, the C$_{60}^-$ beam was exposed 
to a laser pulse from a tunable optical parametric oscillator 
(OPO) laser, with photon energies which were varied between 1.9 
eV and 2.7 eV in steps of 0.1 eV or 0.2 eV.
Pulse energies were kept low, typically a few mJ or lower, to 
ensure single photon absorption conditions. 
Spectra were recorded with laser firing times between 4 ms 
and 35 ms.

\begin{figure}[ht]
\includegraphics[width=0.4\textwidth,angle=0]{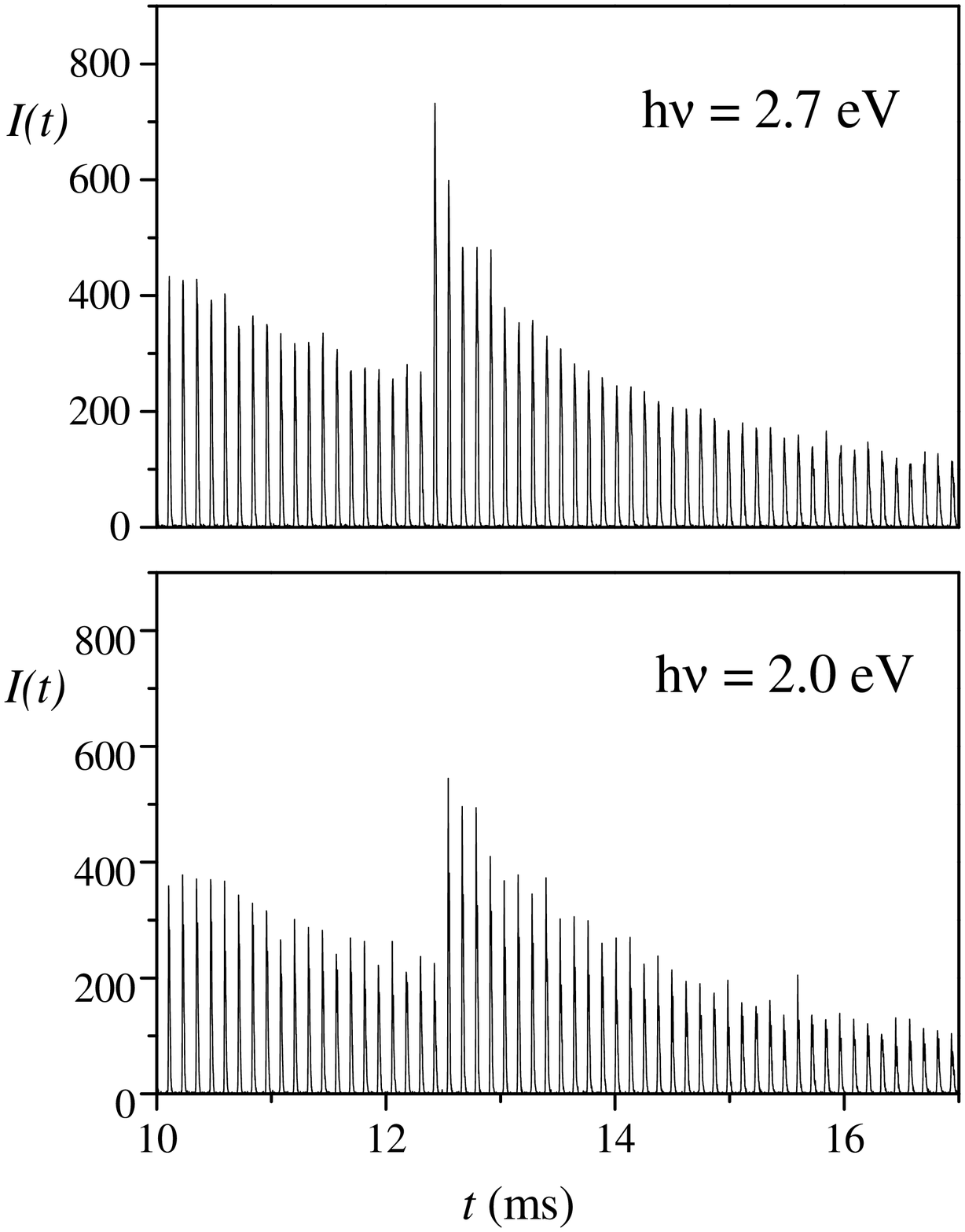}
\begin{centering}
\caption{Two spectra with photo-enhanced decays.\label{twospectra}}
\end{centering}
\end{figure}

Fig.(\ref{twospectra}) shows two example spectra that were 
recorded with laser firing time 12.5 ms and photon energies 
2.0 and 2.7 eV.
Reference spectra without exposure to laser light were 
recorded under identical source and ring conditions, timewise 
interleaving laser-on and laser-off spectra. 
The ion source was found to be very stable, with reproducible 
spontaneous decay rates as a function of time, with variations 
restricted to minor and slow fluctuations in the absolute overall 
ion intensity. 
Such source intensity variations were accounted for by a 
normalization using pre-laser time counts of the laser-on and 
the laser-off spectra. The spontaneous decay showed no
variation with respect to the exponential cutoff caused by 
the radiative cooling, and the decay rate as a function of time 
was in good agreement with the rates previously observed from 
a plasma source \cite{andersen1996}.

The main result of the experiments were the backshifted times 
of the photo-induced decays. As illustrated with a couple of 
examples in \cite{SundenPRL2009}, the photon enhanced signal 
can be represented well by the expression 
\be
R_p(t) \propto \frac{1}{t-t_{las} + t_0},
\ee
where $t$ is the time after production of the ions in the source,
$t_{las}$ is the laser firing time, and $t_0$ the backshifted 
time. 
This simple expression only works for situations where, like here, 
the backshifted time is located in the pure power law sector
before radiative cooling modifies the decay. 
Irrespective of which sector the backshifted time is located,
its interpretation is the same, viz. as the 
reciprocal of the rate constant of the molecule at the energy
$E(t_{las}) + h\nu$, modified with $k_p$ as given above. 

\begin{figure}[ht]
\includegraphics[width=0.3\textwidth,angle=90]{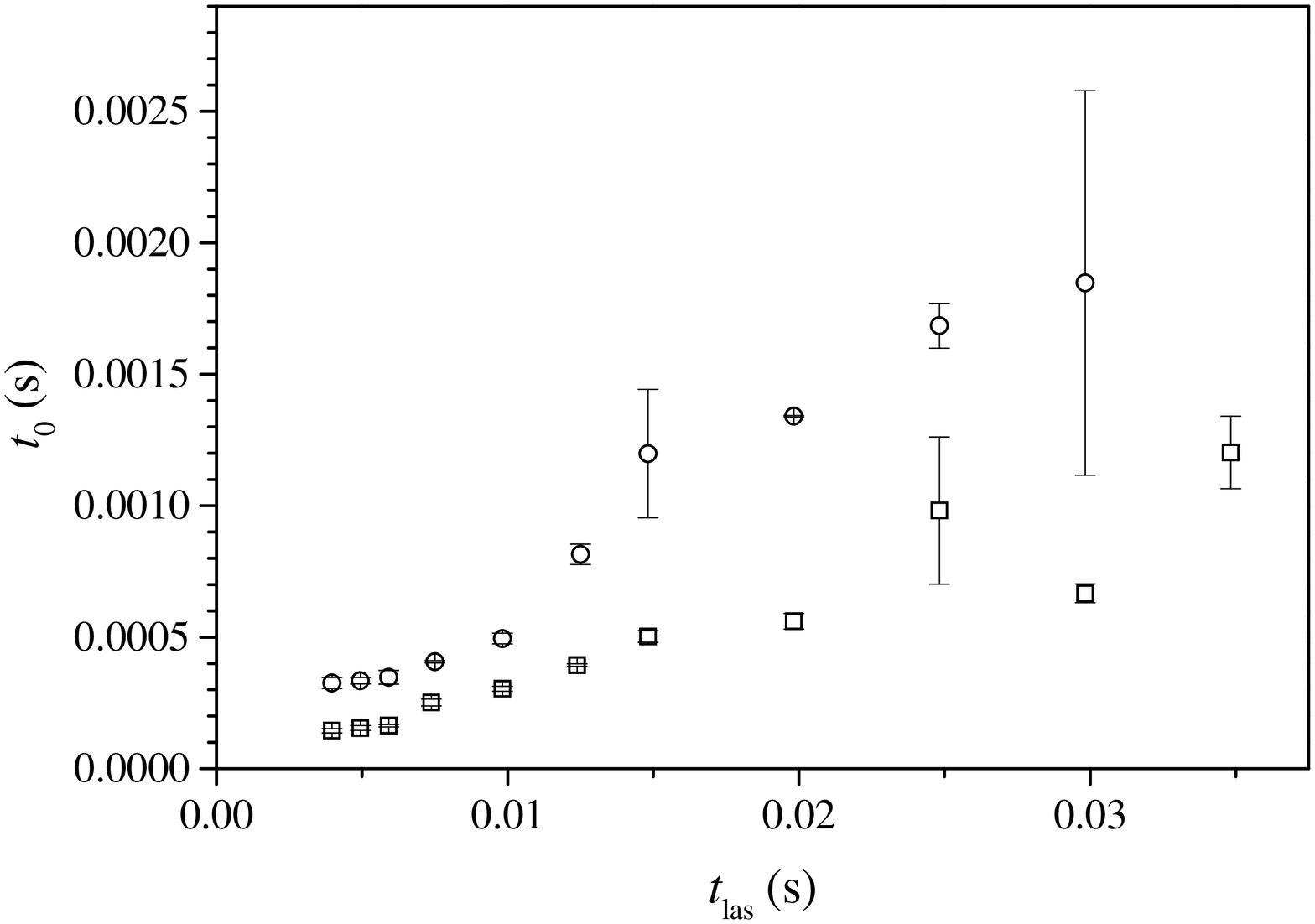}
\begin{centering}
\caption{Traces of $t_0$ as a function of laser firing time 
measured with the photon energies $h\nu =2.0$ eV (circles) 
and 2.7 eV (squares).
The error bars are statistical. 
Some amount of fluctuation beyond statistical are present, 
due to the so-called betatron oscillations, a well-known phenomenon
from ion storage rings.\label{t0fig}}
\end{centering}
\end{figure}

Fig.(\ref{t0fig}) shows examples of the fitted $t_0$ for 
experiments with two different photon energies and a range of 
different laser firing times.

\section{Data analysis}

The data analysis proceeds from the data set comprising associated 
values of laser firing times, $t_{las}$, backshifted times, $t_0$, 
and photon energies, $h\nu$, together with the rate constants for 
these times, $k(E(t))$, derived from the measured decay rates, as 
explained above.

\begin{figure}[ht]
%\vspace{-2cm}
\includegraphics[width=0.3\textwidth,angle=270]{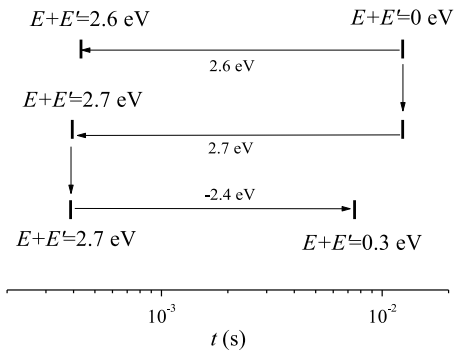}
\begin{centering}
%\vspace{-1cm}
\caption{The computational flow in the calculation linking the relative
energies, illustrated with three laser firing times, two of which are 
identical. 
The energy assignment begins at the top right corner by
choosing this as the zero of energy, and flows in the direction of 
the arrows. 
The points at times around 10 ms correspond to data 
at $t_{las}$, and those at a few hundred $\mu$s are data at the 
corresponding $t_0$'s.
As described in the main text, all points can be
assigned a known rate constant, modulo the value of $k_p$.
\label{C60RCSchematic}}
\end{centering}
\end{figure}

The first part of the analysis is initiated by assigning a zero 
energy arbitrarily to the edge energy, $E(t_{las})$, at some given 
laser time. 
In this case it was chosen to be $t_{las} = 0.00994$ s. 
A few alternative starting points were tried 
without any significant 
change in the result.
The rates at both this times and after absorption of a photon are 
known, as is the difference in energy.
This places all rates measured with the same laser firing time on 
the energy axis with known relative positions. 
Such data for different laser firing times are linked to each other 
when the different $t_0$'s are close, ideally identical, for different 
laser firing times and photon energies. 
The criterion for two $t_0$'s being identical was chosen to be a 
difference of no more than 10 \% in value. 
The computational procedure is illustrated in Fig.(\ref{C60RCSchematic}). 
All 62 measured combinations of laser firing times and photon energies 
were linked to the common energy reference this way. 
The linked energies are independent of the values of $k_p$, but the 
rate constants for each time are not. 
They need to be calculated with Eq.(\ref{kcorr}).

As the value of $k_p$ is not known at this point, curves of the 
thermionic emission rate constant $k(E)$ were 
calculated for different assumed values of $k_p$, varying it from 10 
to 100 s$^{-1}$ in steps of 10 s$^{-1}$. 
For each of these, the logarithmic slope was fitted.
The logarithmic slope takes the form
\be
\label{logderivk}
\frac{\d \ln k}{\d E} = \frac{\phi C_v}{E^2} = 
\frac{\ln \(\omega/\overline{k} \)^2}{\phi C_v},
\ee
where $\overline{k}$ is the logarithmic midpoint of the data 
range for which the derivative is fitted.

The second step in the analysis is taken by considering 
the variation of the rate constants when the laser time 
is changed and the photon energy is kept constant. 
Taking the ratio of the rate constant at the backshifted 
time to the rate constant at the laser firing time one 
gets, with $E_{las}$ denoting the energy edge at the laser 
firing time and using the expression for the 
energy resolved rate constant in Eq.(\ref{kstand})
\begin{eqnarray}
\frac{k(t_0)}{k(t_{las})} &=& \exp \(-\frac{\phi C_v}{E_{las} + h\nu} 
+ \frac{\phi C_v}{E_{las}}\) \\\nonumber
&\approx& 
\exp\( \frac{\phi C_v h\nu}{E_{las}^2} - \frac{\phi (h\nu)^2}{E_{las}^3}\),
\end{eqnarray}
or
\be
\ln\( \frac{k(t_0)}{k(t_{las})}\) \approx 
\frac{\phi C_v h\nu}{E_{las}^2}\(1-\frac{h\nu}{E_{las}}\).
\ee
The left hand side of this expression are  
expressed with the relations between rate constant and time
in Eqs.(\ref{km1},\ref{km2}).
The value of $E_{las}$ can be expressed in terms of the rate 
constant as
\be
k(t_{las}) = \omega \exp\(-\frac{\phi C_v}{E_{las}} \) 
\Rightarrow E_{las} = \frac{\phi C_v}{\ln\( \omega/k(t_{las})\)}.
\ee
Inserting this and taking the square root gives the quasi linear relation
\begin{eqnarray}\label{omegaeq}
\(\ln \(\frac{k(t_0)}{k(t_{las})} \)\)^{1/2} =\hfill\\\nonumber
\(\frac{h\nu}{\phi C_v}\)^{1/2} \( \ln (\omega t_{ref}) - \ln (k(t_{las})t_{ref}) \)
\\\nonumber
\times \(1-\frac{h\nu \ln\(\omega /k(t_{las}) \)}{\phi C_v} \)^{1/2},
\end{eqnarray}
where $t_{ref}$ is a reference time that can conveniently be taken as 1 s.

Repeating this procedure with $t_{las}$ replaced by $t_0$ on the right 
hand side gives a similar result apart from the exchange 
$t_{las} \rightarrow t_0$, and a change of sign on the last term in 
the last bracket. 
Averaging the two and dividing by $\sqrt{h\nu}$ gives
\begin{eqnarray}
\label{omegaeq3}
\frac{1}{\sqrt{h\nu}}\(\ln \(\frac{k(t_0)}{k(t_{las})} \)\)^{1/2}
\approx \hfill \\\nonumber \(\frac{1}{\phi C_v}\)^{1/2} 
\( \ln (\omega t_{ref}) - \frac{1}{2}\ln (k(t_{las})k(t_0)t_{ref}^2) \).
\end{eqnarray}
When evaluating the quality of this approximation, it was compared with 
that of the ratio of rate constants expressed as
\begin{eqnarray}
\ln\(\frac{k(t_0)}{k(t_{las})}\) &=& -\frac{\phi C_v}{E_{las} + h\nu} + 
\frac{\phi C_v}{E_{las}} \\\nonumber
= \frac{\phi C_v}{E_{las}\(E_{las}+ h\nu \)}
&=& \frac{h\nu}{\phi C_v} \ln\(\frac{\omega}{k(t_0)}\) 
\ln\( \frac{\omega}{k(t_{las})} \).
\end{eqnarray}
This is inconvenient for graphical representation, but a test using 
it (not shown) confirms the validity of the above approximation. 

\begin{figure}[ht]
%\vspace{-2cm}
\includegraphics[width=0.3\textwidth,angle=270]{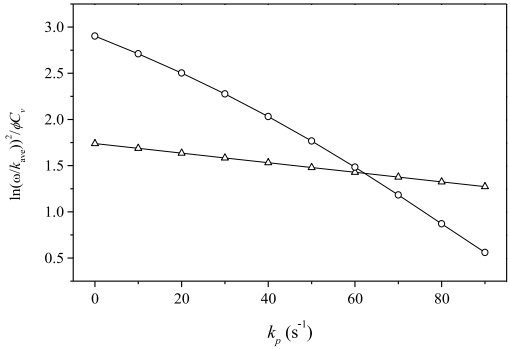}
\begin{centering}
\vspace{-1cm}
\caption{The values of $\ln(\omega \times 1{\rm s})^2/\phi C_v$ vs. $k_p$ 
calculated with Eq.(\ref{logderivk}) (circles) and 
Eq.(\ref{omegaeq3}) (triangles)\label{kpfit}.}
\end{centering}
\end{figure}

\begin{figure}[ht]
\includegraphics[width=0.3\textwidth,angle=270]{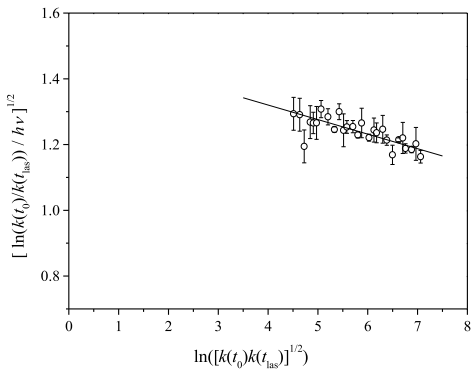}
\begin{centering}
\caption{Plot of the data calculated with Eq.(\ref{omegaeq3})
and $k_p=60$ s$^{-1}$.
The points are grouped in bunches 0.1 wide and the error bars are 
calculated as the statistical average on the mean. 
For a few points where there is only one datum in the bunch, the 
error is set to 0.05.\label{omegafig}}
\end{centering}
\end{figure}

Eq.(\ref{omegaeq3}) defines a straight line. The value of $k_p$ 
enters into the values of $k(t_0)$ and $k(t_{las})$ and hence 
also of the slope and the intercept of the straight line. 
The intercept squared allows a comparison with the value obtained 
with Eq.(\ref{logderivk}) after a correction for the difference between 
time scales used in the factor $\ln(\omega t)$ in the two equations. 
The comparison of the two values is shown in Fig.(\ref{kpfit}) 
vs $k_p$.
Consistency requires identical values for the two curves, yielding
the value $k_p=60$ s$^{-1}$.
This value inserted into Eq.(\ref{omegaeq3}) gives the line in 
Fig.(\ref{omegafig}). 

Another possible contribution to the analysis shown in 
Fig.(\ref{omegafig}) should be mentioned. 
It is obtained by replacing the rate constant at the laser 
firing time with one for a different photon energy, i.e. using 
two different photon energies and hence two different backshifted 
times from the same laser firing time. 
The equation then reads
\begin{eqnarray}
\label{omegaeq4}
\frac{1}{\sqrt{h\nu_1-h\nu_2}}\(\ln \(\frac{k(t_0(1))}{k(t_0(2))} \)\)^{1/2}
~~~~~~~~~~~~~~~~\\\nonumber
\approx \(\frac{1}{\phi C_v}\)^{1/2} 
\( \ln (\omega t_{ref}) - \frac{1}{2}\ln (k(t_0(1))k(t_0(2))t_{ref}^2) \),
\end{eqnarray}
where the arguments (1) and (2) refer to different photon energies at the 
same laser firing time. 
The present data (not shown) are too scattered to provide any strong 
confirmation of the analysis, but are consistent with it.

The parameters of the line in Fig.(\ref{omegafig}) gives the values
\be
\label{fitparam}
\ln (\omega ~1{\rm s}) = 33.8 \pm 6.0,~~~\phi C_v = 510 \pm 180 ~{\rm eV},
\ee
corresponding to a frequency factor of $\omega = 4.9\times 10^{14}$ s$^{-1}$
with a 1-$\sigma$ uncertainty of a factor 400.

\begin{figure}[ht]
\includegraphics[width=0.3\textwidth,angle=270]{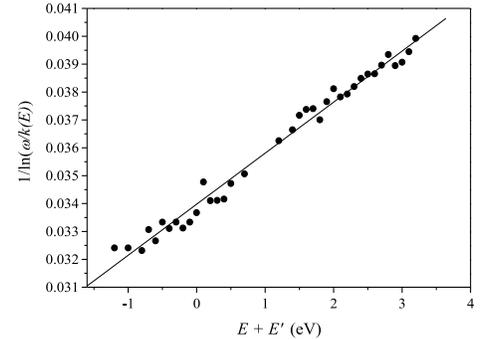}
\begin{centering}
\caption{Plot of Eq.(\ref{EfitEq}) with rate constants calculated with 
the linking procedure explained in the main text, and the value 
$k_p = 60$ s$^{-1}$. 
The line is a straight line fit. 
The parameters of the line give the values $\phi C_v = 546 \pm 12$ eV, 
and $E'= 18.6 \pm 0.4$ eV. The error in $\omega$ is 
not included in 
these two standard deviations.\label{Efit}}
\end{centering}
\end{figure}

\begin{figure}[ht]
\includegraphics[width=0.3\textwidth,angle=270]{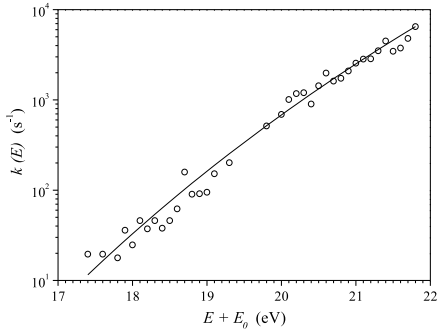}
\begin{centering}
\caption{The thermionic rate constant of C$_{60}^-$ vs. energy.
Error bars can be taken as the average point-to-point fluctuation. 
The full line are the values calculated with the parameters 
$\omega = 4.9\times 10^{14}$ s$^{-1}$, $\phi C_v = 546$ eV, and
$E'=18.6$ eV. The latter is added to the energy. As it includes the 
offset in the caloric curves, this is added as $E_0$.
\label{C60RateConstant}}
\end{centering}
\end{figure}

The above results can be used to verify the procedure by applying them 
to the rate constants found with the linking procedure illustrated in 
Fig.(\ref{C60RCSchematic}). As $k_p$ is known, also these rate constants 
are known, apart from the offset in energy.
The expression for the rate constant is rewritten, reintroducing the 
offset energy $E'$, as 
\be
\label{EfitEq}
\frac{1}{\ln (\omega/k(E))} = \frac{E + E'}{\phi C_v}.
\ee
Using the value of $\omega$ fitted above, the left hand side is 
plotted vs. $E$ in Fig.(\ref{Efit}). 
The expected straight line behavior is observed, and the fitted value 
of $\phi C_v$ is consistent with the previously fitted values, although 
the uncertainty is significant larger than the fitted value indicates.
The rate constant calculated with the two fit parameters from 
Fig.(\ref{Efit}) and the previously determined $\omega$ 
from Eq. (\ref{fitparam}) is shown in 
Fig.(\ref{C60RateConstant}).
Also shown are the measured rate constants from 
Eqs.(\ref{kcorr}), calculated with the large photon energy
parameter value $k_p = 60$ s$^{-1}$, the
experimentally measured rates $R(t)$ and the fitted values of 
the energies based on the measured values of $t_0$, as described 
in detail above.

\section{Discussion}

The analysis has been based on experimental data and the result in 
Fig.(\ref{C60RateConstant}) gives the rate constant from experimental
data alone. The different determinations of the parameters can be 
summarized as a value for the frequency factor of $\omega = 
5 \times 10^{14}$ s$^{-1}$ with an uncertainty of a factor 400;
two values of $\phi C_v$ of which $510 \pm 180$ eV must be considered 
the primary. The second value is consistent 
with this number but is derived 
assuming the above value of the frequency factor.
Finally, the zero of energy has been fitted 
to a value of 18.6 eV. 
The first two parameters in this list have obvious interpretations, 
but also the energy offset contains information about
the reacting species. 

The parameters extracted from the fits differ from the values 
that can potentially be measured
in other experiments because approximating a microcanonical rate 
constant, which is essentially a ratio of level densities with an 
exponential, will generate some finite size corrections.
These corrections were calculated in \cite{hansen2015}.
To leading order in the reciprocal heat capacity they can be 
summarized as
\begin{eqnarray}
C_v = \overline{s} - \frac{\ln (\omega t)^2}{12 \overline{s}},\\
\phi = E_a + E_r - E_t,
\end{eqnarray}
where $\overline{s}$ is the average of the number of thermally activated
oscillators of precursor and product, and the two energies $E_r,E_t$ are
the offsets in the canonical caloric curves for the anion ($E_r$)
and the neutral molecule ($E_t$), defined as:
\be
E = s_iT - E_{i},
\ee 
where $i$ represents either $r$ or $t$ and the $s_r,s_t$ are the 
$s$-values for the reactant (the anion) and the 
transition state (the neutral) as indicated by the subscripts.
$E_a$ is the adiabatic electron affinity with the previously cited 
value of 2.67 or 2.68 eV. 

The correction to the heat capacity is very minor for C$_{60}$, on the 
order on 1 in the units of $k_B$ used, 
and can be ignored here. 
Also the slight variation in the heat capacity due to its temperature 
dependence will be ignored (see the appendix for a 
discussion of this question).

The correction to the activation energy is the most important.
It vanishes for a harmonic oscillator system if the number and 
frequencies are identical in the precursor and product, because for 
harmonic oscillators the offsets are just the sum of their zero 
point energies.
Although the number of oscillators is identical for the anion and the 
neutral molecule and the oscillators can be considered harmonic because 
the degree of excitation is very low, a correction arises because the 
frequencies differ. 

The entire sets of frequencies of the neutral and the anion are 
not known. 
The two anion infrared active modes reported in \cite{KupserPCCP2008} of 
570 cm$^{-1}$ and 1374 cm$^{-1}$ are shifted slightly
relative to the neutral values of 570 cm$^{-1}$ and 1411 cm$^{-1}$ 
\cite{NemesCPL1994}. If the reduction of the highest frequency is used as
the scaling for all frequencies, the corresponding reduction in 
total zero point energy of the anion compared to the neutral is 0.26 eV. 
For this estimate the set of vibrational frequencies of \cite{Menendez} 
was used.
Although these frequencies refer to fullerite and not to gas phase 
molecules, the values should be sufficiently precise 
for this purpose. 
The net result is to reduce the effective activation by 10 \%. At 
the same time the reduced vibrational quantum energies push the 
heat capacity up toward the classical canonical limit of $3N-6$.
The combined effect is therefore less than the 10 \% reduction of the 
activation energy alone. 
As the anion vibrational spectrum is by and large 
unknown, a more accurate estimate of the expected value of 
$\phi C_v$ will not be attempted.

In the definition of an emission temperature which 
is used here, 
some offsets enter the energy content of the decaying 
anions \cite{hansen2015}. 
To a sufficient precision the emission temperature is, in terms of 
the physical excitation energy $E$ equal to 
\be
T_e = \frac{1}{C_v}\(E - \frac{E_a}{2} + E_r \).
\ee
The quantity in the bracket is the energy that appears on the 
abscissa in Fig.(\ref{C60RateConstant}), i.e. the offset energy
$E_0$ is equal to $E_r - E_a/2$. With the reduced frequencies for 
the anion, this amounts to $E_0 = 7.7$ eV. To get the physical 
energy on the abscissa in Fig.(\ref{C60RateConstant}) this number
therefore needs to be subtracted. The rate constant has consequently 
been determined for the range of energies 9.7 to 14.1 eV.

The frequency factor in the simplified expression
for the rate constant used here can be calculated from the 
exact value determined by the expressions given in 
\cite{andersen2002thermionic} with some correction factors which  
can be found in \cite{hansen2015}.
It is not an observable that can be compared with other measurable
quantities, and as a calculation of it involves a number of 
factors with each their uncertainty, a calculation of the its 
value will not be attempted here.

Finally, it is worthwhile to consider the amount of radiative cooling
by low and high energy photon emission. 
The distinction between these two categories is made according to 
whether or not the emission of one photon quenches the electron 
emission channel. 
The large photon energies are defined as \cite{FerrariIRPC2019}
\begin{eqnarray}
\frac{\d \ln k}{\d E} h\nu > 1 \Rightarrow 
h\nu > \frac{\phi C_v}{(\ln(\omega/\overline{k}))^2}\\\nonumber
= \frac{510 {\rm ~eV}}{\ln(4.9\times 10^{14}/400)^2} = 0.66 {~\rm eV}.
\end{eqnarray}
Photons of this magnitude are within thermal reach, 
as can be seen by a calculation of the microcanonical
temperature. For the anion this is $(E+E_0 + E_a/2)/160$, where $E+E_0$ is 
the fitted effective energy content, and $E_a/2$ is the correction 
for the finite heat bath, which can be ignored for photon emission. 
The 160 is the heat capacity. This is slightly less than the 
contribution from all oscillators, which is 174 in the harmonic 
and high temperature limit. 
The calculated effective temperature is then 
0.12 eV for the typical energy of 18 eV.
The phase space of the photon and the quadratic absorption cross 
section \cite{AndersenEPJD2000} makes the total 
emission rate proportional to
the fourth power of the photon energy. 
In terms of the microcanonical temperature:
\be
\label{photonEdist}
k_{photon}(h\nu) \d h\nu \propto 
(h\nu)^4 \frac{\e^{-h\nu/T}}{1- \e^{h\nu/T}}  \d h\nu.
\ee

The total emitted power is bounded from below by 0.66 eV
$\times 60$ s$^{-1} = 40$ eV/s. 
This should be compared with the radiative energy loss of 
approximately 100 eV/s reported in \cite{SundenPRL2009}.
As mentioned, this emitted power refers to the 
radiation emitted as continuous cooling exclusively. 
We can use this value to normalize Eq.(\ref{photonEdist}) and 
find the total emitted power as well as the distribution on 
low and high energy photons. 
Using the photon energies up to 0.66 eV, the 
low energy photon cooling determines the 
constant $c$ as
\be
100 {\rm~ eV/s} = c \int_0^{0.66 {\rm ~eV}} \(h\nu \)^5 
\frac{\e^{-h\nu/T}}{1-\e^{-h\nu/T}} \d h\nu.
\ee
The corresponding high photon energy emission rate constant
is
\be
k_p = c \int_{0.66 {\rm ~eV}}^{\infty} \(h\nu \)^4 
\frac{\e^{-h\nu/T}}{1-\e^{-h\nu/T}} \d h\nu.
\ee   
The value is calculated to 120 s$^{-1}$, i.e. a factor 
2 higher than the fitted value. 
The value decreases to 90 s$^{-1}$ for the temperature 0.11 eV.
Considering that the spectrum in Eq.(\ref{photonEdist}) is 
somewhat schematic, the agreement is reasonable.
In any case, the data suggest that a considerable fraction of 
the radiative energy is emitted as high energy photons.
This is remarkable, both because the systems is as large as it is, 
and because the electron affinity, which acts 
as the activation energy and therefore sets the temperature 
scale, is not particularly large compared with activation 
energies for unimolecular fragmentation, for example. 

\section{Summary and perspectives}

The rate constant for thermal electron emission from C$_{60}^-$ has
been determined over a 4 eV energy range. 
The determination applies a simplified rate constant but 
does not rely on any modeling. 
The experimental input is the set of associated values of 
backshifted times, photon energies and laser firing times 
in a reheating experiment. 
The experiment was performed in a storage ring, which is an ion 
storage device which allows to probe a wide range of times and 
thereby to cover a reasonable internal energy range.

The analysis provided the absolute value and the logarithmic 
derivative of the rate constant with respect to energy, and the 
product of activation energy and heat capacity, together with the
frequency factor for the rate constant. 
The values were found to be in the range of expected and physical
reasonable, although the uncertainties were not negligible.
The main problem of the analysis of the present 
data is the presence of betatron oscillations.
Although these are inherent to the operation of storage rings, 
their magnitude decreases in smaller rings, for simple 
geometrical reasons related to relative detector size 
\cite{MatsumotoNIMB2019}.
The analysis presented here is a proof of principle   
for the method which provides rate constants for large 
systems that are otherwise in practice beyond reach of 
experimental measurement, and the commissioning of 
still smaller storage rings promise the possibility for still
more accurate measurements. 
Other problems may arise with a reduction of the
betatron oscillations. 
Application of the the method to smaller 
molecules/clusters could require that finite heat capacities are 
taken into account. 
A deviation from the straight line behavior seen in Fig.(\ref{approx})
indicate that the finite heat capacity imposes modifications
on the analysis. 
Such effects have been seen in e.g. \cite{MenkPRA2014} for SF$_6^-$.
As the method described in this work is most urgently needed 
for large systems where heat capacities tend to vary smoothly 
with energy over the measured range, such complications are 
not expected to constitute a major drawback.
Another limitation associated with level densities and thermal
properties should be mentioned, viz. the possibility of a 
liquid/solid phase transition or, more correctly, the finite 
size equivalent of this.
This will modify the power law decay as described in 
\cite{HansenRMS2020}, essentially with an increase, in contrast 
to the decrease induced by radiative cooling, and is 
therefore directly observable in the measured spectra of 
spontaneous decays.

\section{Aknowledgements}
My coauthors of \cite{SundenPRL2009} are gratefully acknowledged 
for the productive collaboration that provided 
the experimental material of this paper.

%\bibliographystyle{apsrev} %style file .bst file.
%\bibliography{C60rateconstants} %your .bib file

\section{Appendix: The approximation of the rate constant}

The use of Eq.(\ref{kstand}) requires that parameters 
extracted from the experiments must be corrected 
before they can 
be compared with parameter values from other types of experiments.
The corrections are known \cite{hansen2015} and 
have been applied in the Discussion section.

The energy in the denominator, $E+E'$, is the sum of the true 
thermal energy, $E$, and an offset, $E'$, which is required to 
account for situations where the thermal energy is not simply 
proportional to the temperature.
The offset includes the zero point energy of the harmonic oscillators,
which provide the largest part of the heat capacity of the molecule,
but also accommodates any other thermal offset that may be present 
below $E$, for whatever reasons. In the following this offset will 
be absorbed into the energy and will not appear explicitly.

\begin{figure}[ht]
\includegraphics[width=0.3\textwidth,angle=270]{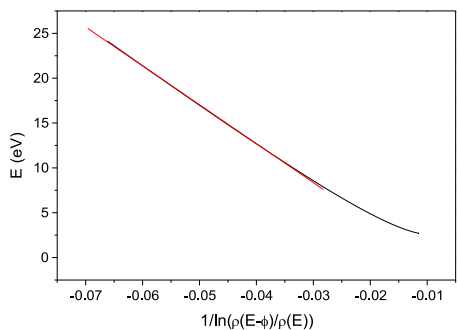}
\begin{centering}
\caption{The test of the approximation of the rate constant by the 
expression in Eq.(\ref{approxeq}).\label{approx}}
\end{centering}
\end{figure}

The main energy dependence of the electron emission rate 
constant is the contribution from the ratio of level densities, 
$\rho$, and the main question therefore concerns the accuracy 
of the approximation 
\be
\label{approxeq}
\frac{\rho(E-\phi)}{\rho(E)}= \exp\(-\frac{\phi C_v}{E+E'} \).
\ee
The quality of this approximation is best seen by plotting 
$E$ vs. $\ln\(\rho(E-\phi)/\rho(E)\)$.
For this purpose the known electron affinity 
and the known frequencies of the molecule in combination 
with the Beyer-Swinehart algorithm are used.
From the rewritten relation
\be
E = \frac{\phi C_v}
{\ln\(\frac{\rho(E-\phi)}{\rho(E)}\)} - E'
\ee
a straight line is expected. 
It is indeed also seen in Fig.(\ref{approx}).
The slope is $434$ eV and the offset gives
$E'=4.64$ eV,both in good agreement with the expected values. 
Importantly, the line is straight to a good approximation.  
The value where the expected abscissa is located is centered 
slightly below -0.04, with an $\pm$ 2 eV range 
at each sides on the ordinate.
This is well in the linear part of the curve.
\end{document}